\documentclass[10pt, conference, letterpaper]{IEEEtran}

\usepackage{amsmath,amsfonts}
\usepackage{graphicx}
\usepackage{textcomp}
\usepackage{mathtools}
\usepackage{xcolor}
\usepackage{blindtext}
\usepackage{algorithm}
\usepackage{algorithmic}
\usepackage{mwe}
\usepackage{subfig}
\usepackage{url}
\newcommand\ignore[1]{}

\newtheorem{definition}{Definition}
\usepackage{multirow}
\usepackage{enumitem}
\title{Location, Location, Location: Simplifying IoT Network Traffic using EDNS}
\title{DNS All Over The Place: Simplifying IoT Network Traffic using EDNS}
\title{It's Not Where You Are, It's Where You Are Registered: IoT Location Impact}
\author{\IEEEauthorblockN{Anat Bremler-Barr}
\IEEEauthorblockA{Reichman University,\\ Israel}
\and
\IEEEauthorblockN{Bar Meyuhas}
\IEEEauthorblockA{Reichman University,\\ Israel }
\and
\IEEEauthorblockN{David Hay}
\IEEEauthorblockA{Hebrew University,\\ Israel}
\and
\IEEEauthorblockN{Shoham Danino}
\IEEEauthorblockA{Reichman University,\\ Israel }

}

\newcommand{\anat}[1]{\textcolor{red}{Anat: #1}}
\newcommand{\barm}[1]{\textcolor{blue}{Bar: #1}}

\begin{document}

\maketitle
\begin{abstract}

This paper investigates how and with whom IoT devices communicate and how their location affects their communication patterns. Specifically, the endpoints an IoT device communicates with can be defined as a small set of domains. To study how the location of the device affects its domain set, we distinguish between the location based on its IP address and the location defined by the user when registering the device. We show, unlike common wisdom, that IP-based location has little to no effect on the set of domains, while the user-defined location changes the set significantly. 


Unlike common approaches to resolving domains to IP addresses at close-by geo-locations (such as anycast), we present a distinctive way to use the ECS field of EDNS to achieve the same differentiation between user-defined locations. Our solution streamlines the network design of IoT manufacturers and makes it easier for security appliances to monitor IoT traffic.

Finally, we show that with one domain for all locations, one can achieve succinct descriptions of the traffic of the IoT device across the globe. We will discuss the implications of such description on security appliances and specifically, on the ones using the  Manufacturer Usage Description (MUD) framework. 

\end{abstract}
\section{Introduction}

The Internet of Things (IoT) is a convergence of several technological advances (e.g., in communications, computer networks, embedded systems, cloud computing, and data science) that dramatically changes the way we use and interact with physical devices. IoT technology—in which various physical devices are connected through a computer network—is present, and in some cases dominate, every sector of our society and day-to-day life, including smart homes, industrial applications, critical infrastructure, and connected cars. IoT devices are ubiquitous already today, however, their number is expected to triple during the 2020s, and be as high as 25.4 billion devices by 2030~\cite{stats_2020}. By 2023, IoT devices are expected to account for 50 percent of all networked devices \cite{cisco_2020}, and therefore, it is imperative to explore their traffic characteristics and which factors affect them. Specifically, this paper focuses on \emph{consumers' IoT devices}, existing today, for example, in smart home deployments. Notice that household penetration worldwide will be 14.2\% in 2022 and is expected to hit 25.0\% by 2026~\cite{statista}. Moreover, consumers' IoT devices are very diverse and include home entertainment, comfort, and lightning, (physical) security, smart appliances, energy management, etc.



\begin{figure}
    \centering
    \includegraphics[width=0.96\linewidth]{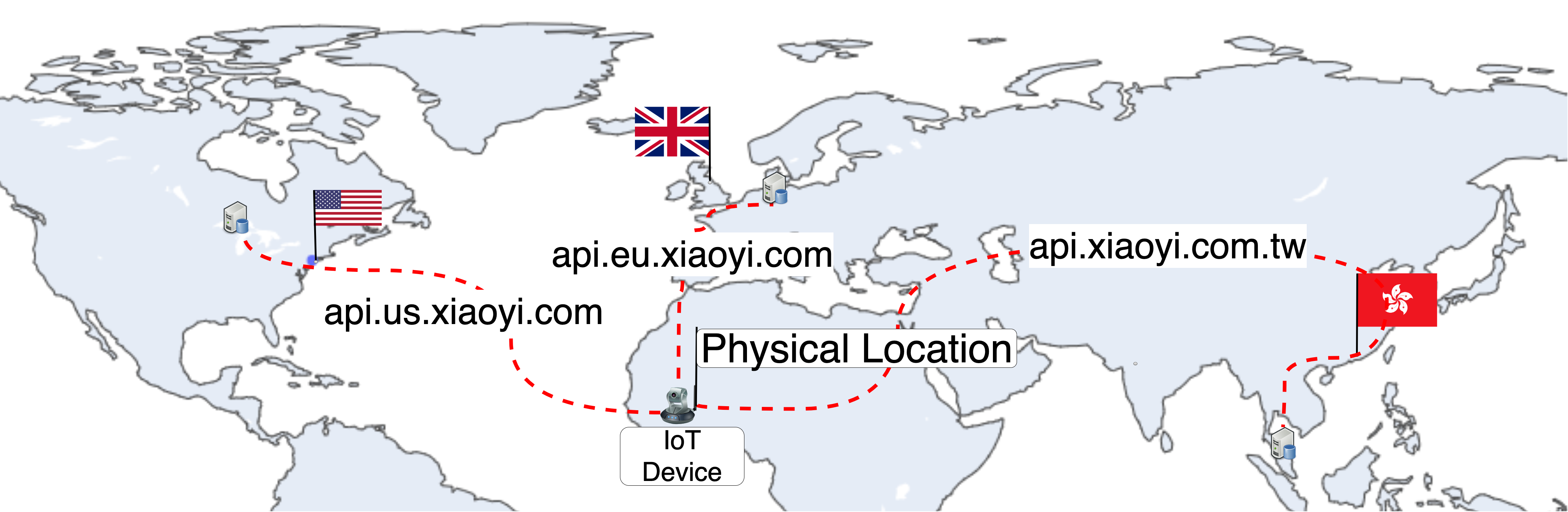}
    \caption[width=0.9\linewidth]{IoT User-Defined Location Impact. Example of a change in domains and servers caused by a change in the user-defined location we registered for, when the IoT device is in the same physical location.}
    \label{fig:first}
\end{figure}

\begin{figure}
\centering
\begin{minipage}{0.86\linewidth}
\centering
\subfloat[Standard DNS Architecture.]{\label{fig:standard_dns_arch}\includegraphics[width=0.86\linewidth]{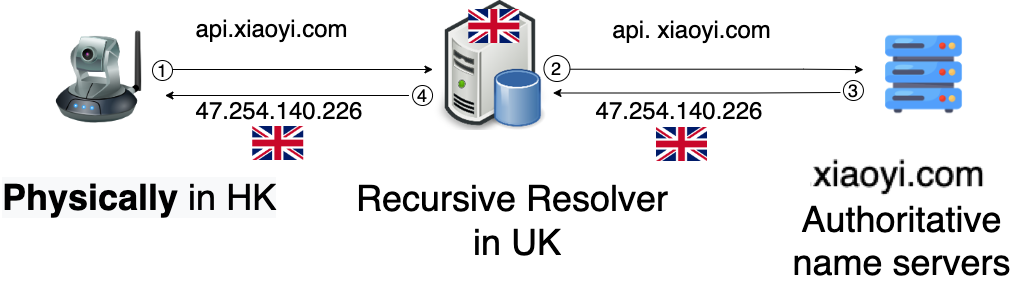}}
\end{minipage}
\vskip 1.2cm
\begin{minipage}{0.86\linewidth}
\centering
\subfloat[ECS Basic Architecture.]{\label{fig:ecs_basic_vs_user_defined_a}\includegraphics[width=0.86\linewidth]{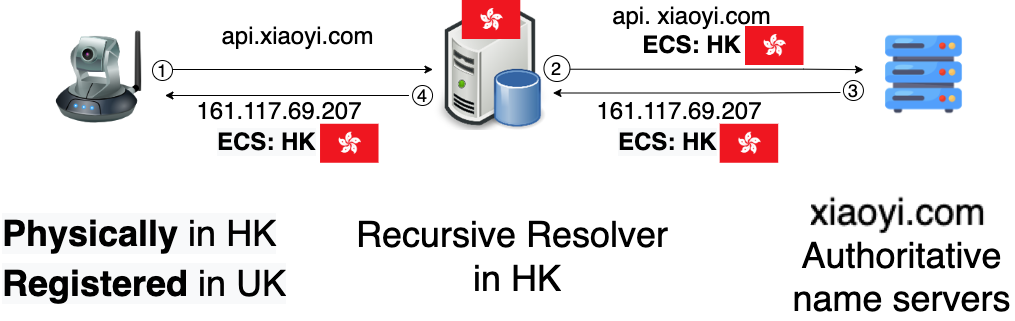}}
\end{minipage} 
\vskip 1.2cm
\begin{minipage}{0.86\linewidth}
\centering
\subfloat[ECS User-Defined Location Architecture.]{\label{fig:ecs_basic_vs_user_defined_b}\includegraphics[width=0.86\linewidth]{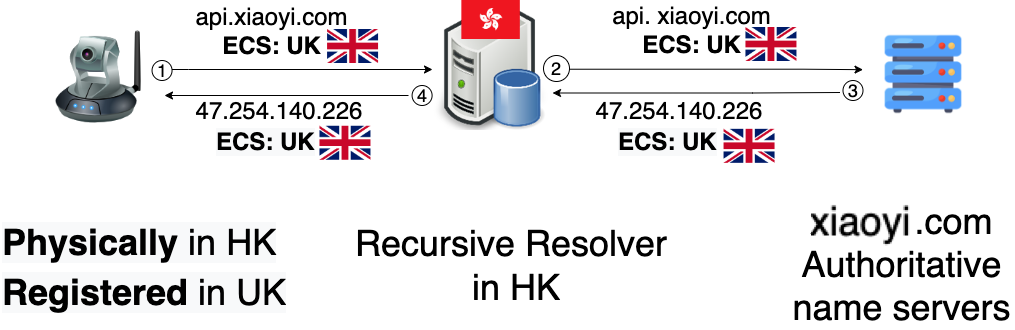}}
\end{minipage}\par\medskip
\caption{
Differences between  resolving a DNS query with Standard DNS, ECS Basic and ECS User-Defined Location Architectures. 
In \ref{fig:standard_dns_arch} the user will receive an IP address related to his resolver IP-based location: \emph{UK}.
In \ref{fig:ecs_basic_vs_user_defined_a}, the \emph{resolver} add the IP-base location of the user in the ECS field and the authoritative reply with the \emph{HK} correlated server.
At \ref{fig:ecs_basic_vs_user_defined_b}, the \emph{device} mentions its user-defined location in the ECS field and the resolver reply with the \emph{UK} correlated server.
(In figure \ref{fig:standard_dns_arch} the resolver in UK and in figures \ref{fig:ecs_basic_vs_user_defined_b} and \ref{fig:ecs_basic_vs_user_defined_a} the resolver in HK)
}

\end{figure}

Unlike general purposes devices (such as computers and smartphones), IoT devices are characterized by the small number of endpoints they access.
In the vast majority of cases in our dataset, which consists of more than 30 devices, a DNS request to a specific domain precedes a connection to an endpoint, thus one can count the number of endpoints by the number of domain names. 
In our dataset, the median number of domains each device connects to is 6 (ranging between 1--57 domains). On the other hand, we show that while the list of domains stays constant in IoT devices, the number of IP addresses they resolved to increases over time.
To be comprehensive, an IoT device's capture should consist of all potential network behaviors, which are sometimes hard to predict. Nevertheless, NIST has defined a list of environmental variables that can influence the network behavior of an IoT device \cite{white_paper_nist} (i.e., internet connection, DNS blocking, human interaction). However, the \emph{location} of the IoT devices is often overlooked. Thus, for example, many studies (e.g, ~\cite{guo2020detecting,mazhar2020characterizing,hu2020toward,hamad2019iot,perdisci2020iotfinder}) that deal with device identification have a datasets captured in a \emph{specific location}. While the derived profiles may be useful to identify the IoT devices in these specific locations, they might lead to biased accuracy results for other locations.

In this paper, we study the impact of the IoT device location on its network characteristic.   We have measured each IoT device at up to 10 locations, and show that, surprisingly, the \emph{user-defined location} (namely, the user-chosen location in the registration process of the user account, required to connect the IoT device to the Internet) is the major factor on the device's network behavior, while the \emph{IP-based location} (namely, the geo-location that corresponds to the device's IP addressed can be retrieved by popular Geo-IP services~\cite{maxmind2019database}) itself has almost no effect. It was very surprising to find that the same device, with the same firmware, behaves differently depending on the user-defined location. This finding affects also other research areas such as device profiling, identification, and security.

Notice that the registration process, and therefore, the user-defined location, is not visible to the network administrator, security appliances, and service providers. Thus, the fact that IoT devices behave differently when changing the user-defined location, makes it significantly harder to monitor, manage, and protect the devices. Moreover, in many cases manufacturers use several domain identifiers to distinguish between (user-defined) device's locations; i.e, two domains were used by the YI Camera in two different user-defined locations: Hong-Kong (\texttt{api.xiaoyi.com.tw}) and United-Kingdom (\texttt{api.eu.xiaoyi.com}, see Figure \ref{fig:first}). Consequently, a security gateway (e.g. firewall), that protects an organizational network with an IoT device, must allow all optional domains for each user-defined location across the globe.


To streamline IoT device security, and make it easier across the globe, we suggest to use one domain name list for all user-defined locations.  We assume that the manufacturers intentionally left the location's decision to the user due to various reasons such as privacy regulation, legal purposes or marketing decisions (e.g., MIUI is a Xiaomi Android fork for specific regions \cite{enwiki:1086751908}). 
Therefore, any proposal must maintain the different functionality required for each user-defined location, implying that the same domain name should be resolved to a different IP address, if the user-defined location is different.

One can suggest using DNS to resolve the same domain to different IP addresses according to the DNS recursive resolver IP address, as present in Figure \ref{fig:standard_dns_arch}, or using the Extended DNS (EDNS) Client Subnet (ECS) field, which allows conveying the user's prefix IP address, as present in Figure \ref{fig:ecs_basic_vs_user_defined_a}. These suggestions would not support the \emph{users decision} of their locations, as indicated, for example, in their registration process. Namely, the recursive resolver will be located in a region corresponding to the IP address of the device, and ECS origin implementation will also send the IP-based location.

We suggest to use the ECS filed of EDNS by \emph{transmitting the user-defined location from the IoT device}. By that authoritative name server receives the user-defined location of the device and is able to respond with the IP address corresponding to the IoT user-defined location, as presented in Figure \ref{fig:ecs_basic_vs_user_defined_b}. 
To the extent of our knowledge, we are the first to propose adding a value at the ECS field at the end-point device.

Using the Open Observatory of Network Interference (OONI) \cite{radu2020consolidation}) dataset, we have analyzed 8 major public DNS providers and ISPs around the world, using the RIPE ATLAS machines. We have found that the ECS field is supported by 7 out of the 8 most popular DNS providers, accounting for 86.20\% of the market of these providers. In addition, we show that the ECS field is being forwarded in all the resolvers we have checked. 




Finally, we show the properties of the set of domain names directly affect IoT security tools that use this set to detect malicious traffic. Specifically, the IETF Manufacturer Usage Description (MUD) framework generates an allow list according to this set. We show how by using ECS, the number of entries in this allow list (namely, the number of domains) is significantly reduced and the entire MUD's allow list management and deployment processes become more streamlined.

\section{IoT Traffic Analysis}

In this section, we present our findings regarding two important properties of consumer IoT traffic: what is the best way to define the endpoints with whom a device communicates and the effect of the  \emph{location} and \emph{registration place} on these endpoints. 

\subsection{The Dataset}
Our findings are supported by a survey we have conducted on IoT network traffic captured from the router in our lab, and log files from Ren et al.~\cite{ren2019information}. Our captures comprise 31 different IoT devices (e.g., plugs, cameras, bulbs, and so on) that are located in up to 14 countries and use all of their device functionalities. 
The entire dataset is publicly available in~\cite{dataset_github1}.  Table~\ref{tab:devices_locations} lists the different devices, their locations, and performed actions. The devices belong to the following categories : cameras, smart hubs, home automation , TVs (actual TVs and TV dongles),audio ,and appliances. 

\begin{table}
\centering
\resizebox{\columnwidth}{!}{%
\begin{tabular}{|c|c|c|c|}
\hline
\textbf{Device}                                                                                                                                                     & \textbf{Functionality}                                                                                                                    & \textbf{Locations}                                                                                                                          & \textbf{Type}                                                              \\ \hline
Sousvide                                                                                                                                                            & \begin{tabular}[c]{@{}c@{}}Power on,\\ Power off\end{tabular}                                                                             & UK, US                                                                                                                                      & Appliances                                                                 \\ \hline
\begin{tabular}[c]{@{}c@{}}Amazon \\ echoplus \\ Amazon \\ echodot\\ Amazon \\ echospot \end{tabular}                                                                                                           & \multirow{2}{*}{\begin{tabular}[c]{@{}c@{}}Power on,\\ Power off,\\  Voice commands \end{tabular}}                        & \multirow{2}{*}{UK, US}                                                                                                                     & \multirow{2}{*}{Audio}                                                     \\ \cline{1-1}
\begin{tabular}[c]{@{}c@{}}Google \\ home mini\end{tabular}                                                                                                          &                                                                                                                                           &                                                                                                                                             &                                                                            \\ \hline
\begin{tabular}[c]{@{}c@{}}Blink camera\\ Wansview cam\\ Ring doorbell\\ Yi camera\end{tabular}                                                                     & \multirow{3}{*}{\begin{tabular}[c]{@{}c@{}}Power on,\\ Power off,\\ Movement,\\ Record video,\\ Take picture\end{tabular}}                & UK, US                                                                                                                                      & \multirow{3}{*}{camera}                                                    \\ \cline{1-1} \cline{3-3}
Yi camera                                                                                                                                                           &                                                                                                                                           & \begin{tabular}[c]{@{}c@{}}UK, US,\\ Hong Kong,\\ Australia,\\ France,\\ Germany,\\ Mexico, India\\ Russia\end{tabular}                       &                                                                            \\ \cline{1-1} \cline{3-3}
Xiaomi camera                                                                                                                                                       &                                                                                                                                           & \begin{tabular}[c]{@{}c@{}}US,\\ China,\\ Israel\end{tabular}                                                                               &                                                                            \\ \hline
\begin{tabular}[c]{@{}c@{}}xiaomi cleaner\\ T-wemo plug\\ Tplink plug\\ Tplink bulb\\ Nest T-stat\\ Magichome strip\end{tabular}                                    & \multirow{3}{*}{\begin{tabular}[c]{@{}c@{}}Power on,\\ Power off,\\ Change brightness,\\ Change temperature,\\ Change color\end{tabular}} & UK, US                                                                                                                                      & \multirow{3}{*}{\begin{tabular}[c]{@{}c@{}}Home\\ Auto- \\mation\end{tabular}} \\ \cline{1-1} \cline{3-3}
\begin{tabular}[c]{@{}c@{}}Tplink plug\\ Lifx \\ light bulb\end{tabular}                                                                                               &                                                                                                                                           & \begin{tabular}[c]{@{}c@{}} UK, US, \\ Hong Kong   \end{tabular}                                                                                                                        &                                                                            \\ \cline{1-1} \cline{3-3}
\begin{tabular}[c]{@{}c@{}}Xiaomi\\ light bulb\end{tabular}                                                                                                         &                                                                                                                                           & \begin{tabular}[c]{@{}c@{}}Spain,\\ Russia,\\ India,\\ Brazil,\\ Australia,\\ Antarctica,\\ Argentina,\\ UK,\\ US,\\ Hong Kong\end{tabular} &                                                                            \\ \hline
\begin{tabular}[c]{@{}c@{}}Samsung\\ smart-things\\ hub \\ Lightify hub \\ Philips hub\\ Blink hub \\ Xiaomi-hub \\ Sengled-hub\\ Insteon-hub\end{tabular} & \begin{tabular}[c]{@{}c@{}}Power on,\\ Power off,\\ Change brightness,\\ Change color,\\ Change temperature\end{tabular}                  & UK, US                                                                                                                                      & Smart Hub                                                                  \\ \hline
\begin{tabular}[c]{@{}c@{}}Appletv \\ Roku-tv \\ Firetv\\ Samsung tv\end{tabular}                                                                                   & \begin{tabular}[c]{@{}c@{}}Power on,\\ Power off,\\ Voice commands,\\ Change Volume\end{tabular}                                          & UK, US                                                                                                                                      & TV                                                                         \\ \hline
\end{tabular}%
}
\caption{IoT devices in our dataset, totally 31 devices. Note to mention that two devices appears both in our dataset and in \cite{ren2019information}, we present them as two separate devices since they have different firmware versions and different network captures.}
\label{tab:devices_locations}
\end{table}

\subsection{Domain names and IP addresses}
 As many IoT devices communicate with services (either dedicated services or generic services) in the cloud, a common practice when designing IoT devices is to refer to these services by \emph{domain names}~\cite{rfc6991} rather by \emph{IP addresses}. Thus, before the (first) connection to an endpoint, the device issues a \emph{DNS request} with the corresponding domain name. Such DNS request is typically issued before any new connection.
 
\begin{figure}[tb]
    \centering
    \includegraphics[width=0.9\linewidth]{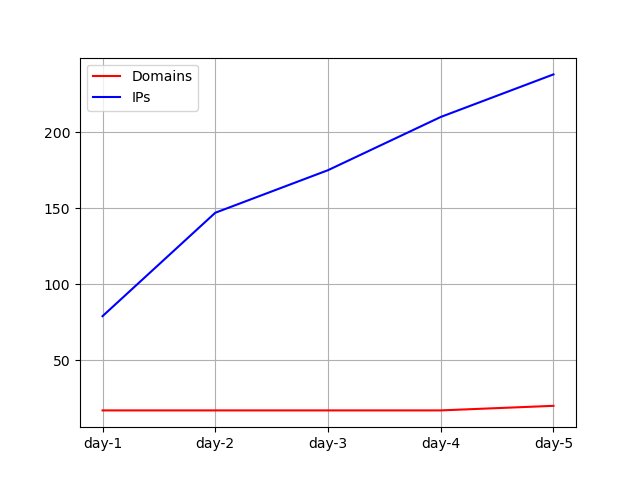}
    \caption{Cumulative number of unique IP addresses and domain names of the Amazon Echo Plus, captured in UK. While the number of domain names is constant along the days, the number of IP addresses increases every day.}
    \label{fig:tp_link_IPs_Domains_along_days}
\end{figure}

 Unlike general-purpose devices, most IoT devices communicate with a pre-defined set of domains. Yet, these domain names may be resolved to different IP addresses over time. 
Fig.~\ref{fig:tp_link_IPs_Domains_along_days} compares the number of domain names (extracted from DNS requests) and the number of different IP addresses used by a specific device, an Amazon Echo Plus. The set of unique domains does not change after one day of observation, while the number of unique IP addresses continues to rise. This discrepancy between the trends happens as IoT devices use cloud services, that have a fixed domain name but dynamic IP address, especially when services are provided as \emph{managed service} by the cloud vendor. Amazon AWS and Microsoft Azure, for example, publish periodically their IP address ranges for their cloud services,  \cite{aws_services,microsoft2022}. Finally, we note that each IP address (besides the DNS server itself) used by the device had a preceding DNS request.  

Thus, in the rest of the paper we refer to the \emph{domain names set} of each device, captured by the following definition:

\begin{definition}
\label{def:domainset}
For an IoT device $d$, let $\mathcal{D}(d)$ be the set of domain names, extracted from DNS requests of device $d$, and $\mathcal{D}(d)|_{(t_0,t_1)}\subseteq\mathcal{D}(d)$ is the set of domain names of device $d$, captured during time interval $(t_0,t_1)$.  A set of domain names \emph{stabilizes} at time $t'$ if for every $t\geq t'$, 
$\mathcal{D}(d)|_{(t_0,t)}=\mathcal{D}(d)$, where $t_0$ is the beginning of the trace.
\end{definition}

In our dataset, the set of the domain names of most devices has not changed within a day of observation. We assume that this is due to the fact that some controlled operations under laboratory conditions (e.g. trying all the options) were done of the IoT devices, and therefore, even shorter traces contain relatively rare events. For some devices, the set of domain names has changed throughout the observation period, this includes, for example, smart TVs that can practically connect for any content provider or even browse the Internet. Some sets have changed due to using pools of domain names (e.g., for load balancing purposes), where each connection to the pool is done through a different domain name within the pool (e.g., to \texttt{\small czfe10.front01.iad01.production.nest.com}, \texttt{\small czfe11.front01.iad01.production.nest.com}, etc. 
We therefore treat these domain names as one, represented by the corresponding regular expression (e.g., \texttt{\small czfe[10-120].front01.iad01.production.nest.com}).



We note that our paper focuses on the domain names, however, IoT devices can connect to additional IP addresses without previously resolving the address from the DNS query. These IP addresses can be either fixed IP addresses (coded in the firmware) or dynamic addresses of either other smartphones on the same LAN (which are very common in IoT devices~\cite{bremlerMUDirect}) or smart-phone that connect to the IoT device using some P2P technology (such as  port-forwarding~\cite{PortForwarding}, UPnP dynamic port forwarding~\cite{rfc6970}, or Hole-Punching techniques~\cite{ford2005peer}, such as STUN/ICE protocols~\cite{STUN_RFC, ICE_rfc8445})

We note that the fact that well-defined domain names sets have motivated security appliance to protect IoT devices using allow-lists, specifying all domain names in $\mathcal{D}$. An important example is the MUD standard which will be discussed in more detail in Section~\ref{sec:MUD}.


\ignore{
\begin{figure}
    \centering
    \includegraphics[width=0.9\linewidth]{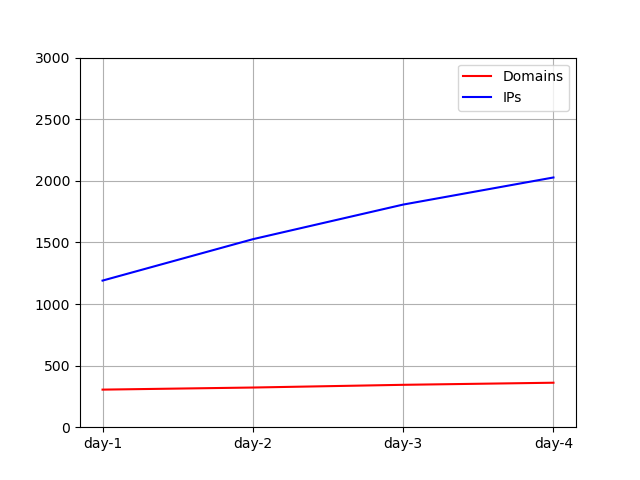}
    \caption{Cumulative number of unique IP addresses and domain names, across 32 devices that captured for up to 10 days. The graph presents the last 4-days.}
    \label{fig:IPs_Domains_along_days}
\end{figure}
}

\subsection{The Impact of Device Location}

Cloud networks, as well as content-delivery networks, typically take into account the geographic locations of their clients, when choosing the right service instance for them (e.g., the closest one, so that delay will be minimized).
In some examples, the geo-location determines whether a client is permitted to use a service or the type of service it gets.

Next, we study how the location of the device affects its domain name set. 

For each IoT device, we distinguish between two types of locations. First, the \emph{IP-based location} of the device is the geo-location of the (external) IP address it uses. For example, the UK is the IP-based location of IoT devices that connect to the Internet through a British ISP. On the other hand, \emph{the user-defined} location is the location the user of the IoT device chose when registering the device (e.g., through a profile). For example, the US is the user-defined location of IoT devices that their users are registered in the US, even if they connect through a British ISP.

To study the effects of both location types on the domain name sets, we let $\mathcal{D}(d,\ell,\ell')$ be the domain name set of device $d$, whose IP-based location is $\ell$ and user-defined location is $\ell'$. We define the pair-wise similarity measure between two locations as the Jaccard similarity coefficient \cite{enwiki:1076414544}  (namely, Intersection over Union) of their domain name sets. This is captured by the following two definitions: 
\begin{definition}
For a device $d$, IP-based location $\ell$ and two user-defined locations  $\ell'$ and $\ell''$ , the \emph{user defined similarity}, denoted $\mbox{uds}(d,\ell,\ell',\ell'')$, is 
\[\mbox{uds}(d,\ell,\ell',\ell'')=\frac{|\mathcal{D}(d,\ell,\ell') \cap \mathcal{D}(d,\ell,\ell'')|}{|\mathcal{D}(d,\ell,\ell') \cup \mathcal{D}(d,\ell,\ell'')|};\] namely, the Jaccard similarity coefficient of the domain name set when switching the user-defined locations from location $\ell'$ to $\ell''$, while the IP-based location remains in $\ell$. 
\end{definition} 
\begin{definition}
For a device $d$, user-defined locations $\ell$, and two IP-based locations $\ell'$ and  $\ell''$, the \emph{IP-based similarity}, denoted $\mbox{ipbs}(d,\ell,\ell',\ell'')$, is 
\[\mbox{ipbs}(d,\ell,\ell',\ell'')=\frac{|\mathcal{D}(d,\ell',\ell) \cap \mathcal{D}(d,\ell'',\ell)|}{|\mathcal{D}(d,\ell',\ell) \cup \mathcal{D}(d,\ell'',\ell)|};\] namely, the Jaccard similarity coefficient of the domain name set when switching the IP-based locations from location $\ell'$ to $\ell''$,  while the user-defined location remains in $\ell$. 
\end{definition} 

\begin{figure}[tb]
\centering\includegraphics[width=0.9\linewidth]{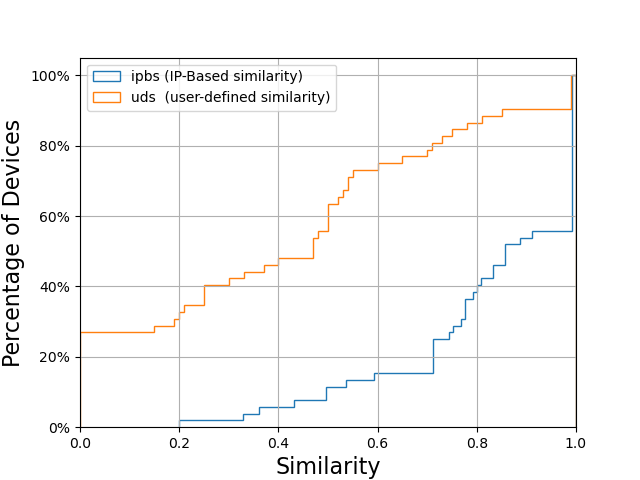}
\captionsetup{width=\linewidth}
\caption{The Cumulative Distribution Function (CDF) of IP-based and user-defined similarities for 26 devices in our dataset, recorded and registered in the US and the UK. 
}
\label{fig:vpn_vs_reg}
\end{figure}

\begin{figure}
    \centering
    \includegraphics[width=\linewidth]{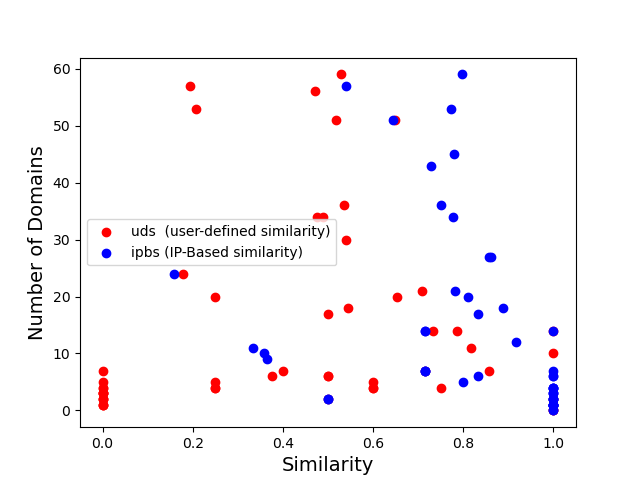}
    \caption{Scatter graph of the user-defined and IP-based similarities of devices in our dataset, by the maximum number of domain names they use, across the two-locations. }
    \label{fig:scatter}
\end{figure}

Figures~ \ref{fig:vpn_vs_reg} and \ref{fig:scatter} present our comparison between the IP-based and location similarities. The study was done based on 26 devices with IP-based and user-defined locations in the US and UK (namely, for every device $d$ we have two values for user-defined similarities: $\mbox{uds}(d,\mbox{US},\mbox{US},\mbox{UK})$ and $\mbox{uds}(d,\mbox{UK},\mbox{US},\mbox{UK})$,  as well as two values for IP-based similarities:  $\mbox{ipbs}(d,\mbox{US},\mbox{US},\mbox{UK})$ and $\mbox{ipbs}(d,\mbox{UK},\mbox{US},\mbox{UK})$. We note that changing the IP-based location was done by setting a VPN tunnel from one country to another and sending the traffic through that tunnel. 
Our finding shows that surprisingly the \emph{user-defined location} of the device has more significant effects than the {IP-based location}. 44\% of the devices do not experience any difference while changing their IP-based location, while 90\% of the devices differ when changing their user-defined location. As seen in Figure~\ref{fig:vpn_vs_reg}, even when there are changes in the domain name set, they are more minor, when considered IP-based similarity. Figure~\ref{fig:scatter} shows that when the size of the domain name set is small, the sets tend to be either disjoint (for user-defined locations) or equal (for IP-based locations).

\begin{figure}
    \centering
    \includegraphics[width=0.8\linewidth]{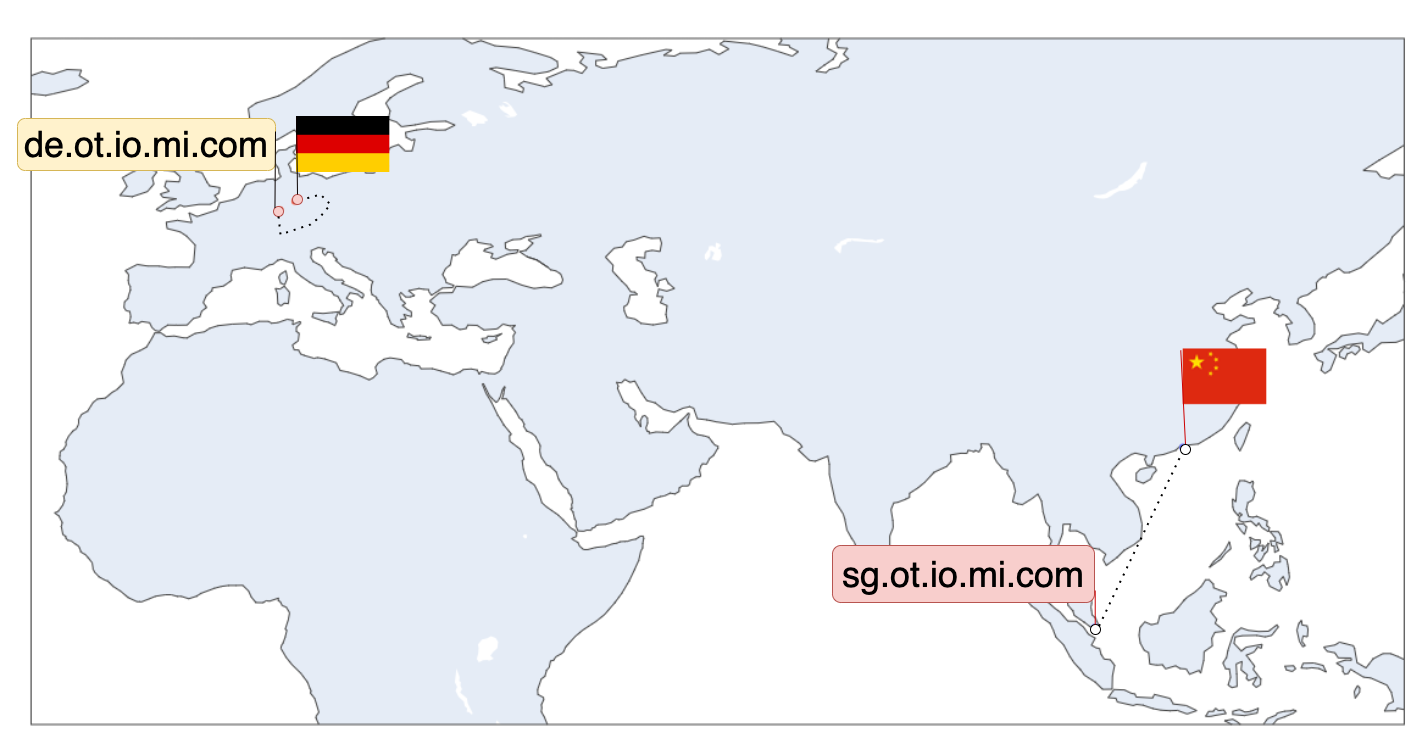}
    \caption[width=\linewidth]{Xiaomi Camera connects to two different domains when registered in China (\texttt{sg.ot.io.mi.com}), and in Germany (\texttt{de.ot.io.mi.com}). The domains were resolved to two different IP addresses in different geographical locations, correlated to the user-defined locations.}
    \label{fig:xiaomi_germany_china}
\end{figure}

\begin{figure*}[tb]
\begin{minipage}{.45\linewidth}
\centering
\subfloat[Xiaomi light bulb]{\label{fig:regional_eq_metric_xiaomi}\includegraphics[width=0.96\linewidth]{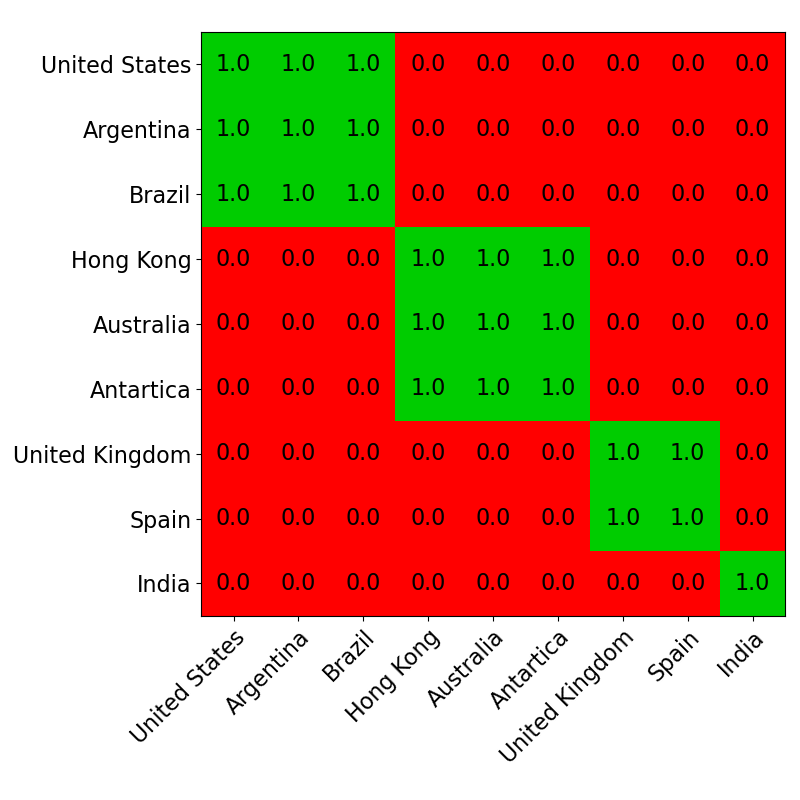}}
\end{minipage}%
\quad\quad
\begin{minipage}{.45\linewidth}
\centering
\subfloat[Yi camera]{\label{fig:regional_eq_metric_yi}    \includegraphics[width=0.96\linewidth]{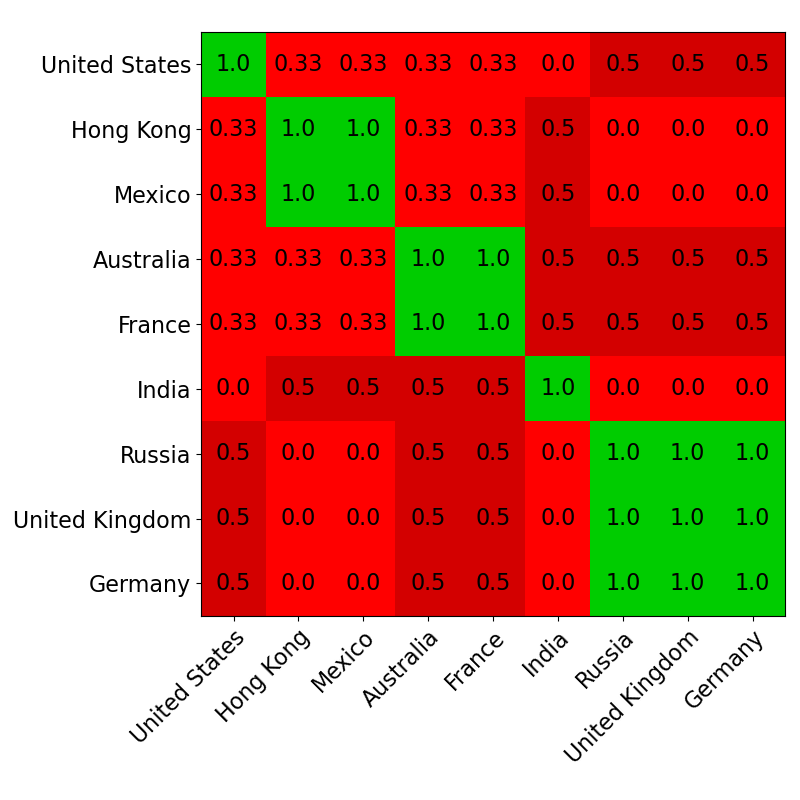}
}
\end{minipage}\par\medskip

\caption{Heatmaps of Used-defined similarities for two IoT devices: (a) Xiaomi light bulb and (b) Yi camera. All measurement were done with IP-based location in a third country (identical for all measurements and different for all user-defined location) and in up to two different user-defined location. For example the value of top-right cell in Fig. ~\ref{fig:regional_eq_metric_xiaomi}
is 
$\mbox{uds}(\mbox{light bulb},X,\mbox{US},\mbox{India})$. The exact IP-based location $X$ is omitted due to the double-blind review process.} 
\label{fig:heatmaps}
\end{figure*}

We further used our dataset to understand what are the differences across user-defined locations. We found that 80\% of the devices use sub-domains to differ user-defined locations, an example is presented in Figure \ref{fig:xiaomi_germany_china}. Nonetheless, 9\% of the devices in the dataset exhibited a difference in the top-level domain (TLD), see an example in Figure \ref{fig:first}. All the different domains across user-defined locations were resolved to different IP addresses, which were geographically close-by to the \emph{user-defined} location. 

In Figure \ref{fig:heatmaps}, we present heat maps of two devices as measured in up to 10 user-defined locations and the same IP-based location. It can easily be observed that each of the presented devices supports several user-defined \emph{regions}, each with different domain names. In many cases, the domain names sets across regions are disjoint. Furthermore,  the heat maps show that sometimes several registration options (and therefore, several user-defined locations) map to the same region. 

Finally, we note that using the same domain name in two different IP-based locations does not necessarily imply that both locations will resolve the domain name to the same IP address. DNS built-in mechanisms, such as anycast and ECS, often resolve a domain name to close-by addresses (in the sense, IP-based locations)\footnote{To demonstrate the change in the IP address responses of a single authoritative name server, we have experimented on the domain \texttt{ebay.com}. We have sent a DNS request to this domain from 147 different machines in 24 different countries. We have not received any identical IP addresses for machines in different countries (in a few cases for different machines in the same country we got the same IP answer), and in 21 countries out of 24 (all except Russia, China, and Ireland) the IP address of the services was in the geographical area of the country from which the query was sent.}. However, these mechanisms take into account only the IP-based location of the device and not its user-defined ones, forcing IoT vendors to use a different domain name for each user-defined location. In the next section, we will show how to circumvent this problem by allowing DNS's ECS mechanism to capture also user-defined locations.

\ignore{
\anat{ Where is the paragraph with all the reasons for registration based decision: Marketing... cloud ... where are the references ??? } 
\barm{i dont think that this is the place for that. This focuses on \textbf{how} the traffic is changed across location and not why.. the why is in the intro, ecs and demonstration}
}


\section{Extended DNS Client Subnet (ECS)}
\subsection{DNS \& ECS Background}
The Domain Name System (DNS) acts as the Internet phonebook and is responsible to translate domain names to IP addresses.

At an abstract level, the DNS system has two parts, each of which is a large,  highly-distributed system: a hierarchical and dynamic database of \emph{authoritative} name servers storing the DNS data of the domain within the authoritative name server zone, and a large number of client-facing \emph{resolvers}, located either locally at the Internet Service Providers (ISPs) and local organizations, or as public services (e.g.,  Google's 8.8.8.8). (Recursive) resolvers walk through the hierarchical structure of authoritative servers to retrieve the domain name resolutions to IP addresses, then return the result to the client, and store the result in the resolver's cache, to reduce the load (in terms of number of DNS requests) of the
authoritative server.

Traditionally, when a domain name is mapped to different IP addresses, the authoritative server returns the IP address closest to the recursive resolver which issued the DNS request to the authoritative server. 
(see Figure \ref{fig:standard_dns_arch}). If the resolver resides within the ISP, the location of the resolver is a good approximation of the end-user location. Thus, this mechanism reduces the distance, and therefore also the latency, between the end-user and the server at the IP address it has requested.

While most resolvers were located in ISPs in the past, nowadays there is an increasing number of open public DNS services. For such services, this approximation is no longer accurate since the resolvers are not necessarily close to the user~\cite{de2019passive}.

Therefore, all public resolvers use an alternative method: the \emph{anycast approach}. In this approach, the resolvers have an anycast IP address that maps to different servers across the globe (e.g., Google's Public DNS is in 8.8.8.8, which is an anycast address that corresponds to 338 different servers). Thus, the recursive resolver issuing the requests to the authoritative server should be close-by to the end-users, implying that the traditional method provides a good approximation and choose an IP address close-by to the end-user.  

However, recently, it was shown that public resolvers users are not necessarily using a resolver that is close to them \cite{google_open_dns_ips}.
Hence, the authoritative started using an Anycast approach, and as a result, the end-users navigated to servers by BGP 
\cite{de2019passive}. 
This failure to locate precisely the end-user's location led to introduction EDNS-Client-Subnet (ECS) solution in RFC 7871\cite{rfc7871}.
EDNS ECS is an extension to the DNS which allows recursive resolvers to convey to authoritative name servers a prefix of the IP address of the client requesting resolution service from the recursive resolver, a demonstration is presented in Figure \ref{fig:ecs_basic_vs_user_defined_a}. 

\subsection{User-defined Location Using ECS}
ECS is intended to help speed up the data delivery by giving the accurate location of the user to the authoritative name server, even when the resolver is not close to the user. In this mechanism, the resolver adds the end-user IP address prefix to the DNS request it sends the authoritative.
In this paper, we suggest using the ECS to support \emph{user-defined locations}. The idea is that the \emph{IoT device}, and not the resolver, will add the ECS to define the location that the user wants to register to.
To the extent of our knowledge, we are the first to propose a solution to the user-defined location, and the first to use ECS by the end device (namely, the IoT device) and the resolver. Moreover, the use of a ECS with a value which is  uncorrelated to the geo-location of the device but with a user-defined value was not proposed yet. We note that according to  RFC 7871~\cite{rfc7871}, it is allowed that a stub-resolver (the IoT device here can act the role of stub-resolver) will add the ECS, even though the general use-case is added by the open resolver.

The minimal requirements from the IoT, resolver, and the authoritative server to support the user-defined location, are as follows:
\begin{enumerate}
    \item 
 The IoT device firmware needs to map each user-defined location, where the user can register, to an equivalent network prefix IP address, and add the prefix to the ECS in the DNS query. Note that currently, each device firmware is built with a list of supported locations and its corresponding domains. We suggest simplifying it and replacing it with a prefix list.
\item The resolver should forward packets with ECS to the authoritative name servers and not modify them.
In the next section, we will show that all the DNS resolvers we tested that enable ECS, also perform forward ECS. In addition, while measuring the adoption rate of ECS, we show that there is high support for ECS. Moreover, if the resolver, does not support ECS, this can easily be overcome by configuring the DNS resolver of the IoT with one of the many open resolvers that support ECS.
\item The authoritative server needs to be configured to enable decision answers based on the ECS, exactly in the same way as in the regular use of ECS with geo-location ECS. All the authoritative server software we have checked are supporting ECS configuration in their latest versions (such as BIND, NSD, BIG-IP, Knot, and Unbound).
\end{enumerate}
\subsection{ECS measurements}
\begin{figure}
    \centering
    \includegraphics[width=0.96\linewidth]{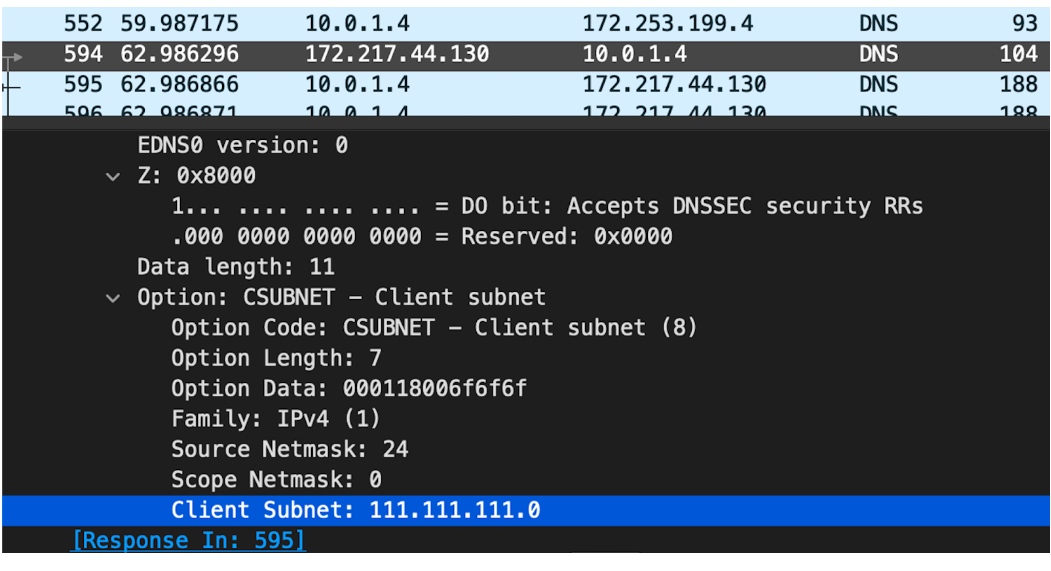}
    \caption[width=0.9\linewidth]{Google Resolver ECS forwarding experiment. We got the answer to our DNS request with the ECS we sent from the client and Google's open DNS resolver forward it to our authoritative name server without modifications.}
    \label{fig:google_ecs_example}
\end{figure}

To check if the open resolvers support ECS, we configured our client to those open resolvers, (listed in Table \ref{table_ecs_enable_open_dns}),  we sent a DNS request with ECS: 111.111.111.0/ 24 and checked if we received the same ECS as a result.
In Figure \ref{fig:google_ecs_example}, we present an example when we subscribed to Google's open resolver (8.8.8.8). Google forward our ECS and we got an answer with the initial ECS we sent
\footnote{8.8.8.8 is the Google Anycast IP, and the Google Resolver IP from which we received an answer is 172.217.44.130, more information is available in \cite{google_dev}}.

In our experiments, we examined two things: that the resolver accepts the ECS we sent and it does not change the ECS value to the IP address of the machine from which we sent it - we call it {ECS forward}. We tested all the resolvers that indicate themselves as ECS enabled and checked if there are also ECS forwarding.  
From the Open Observatory of Network Interference dataset, we retrieve the market share in 2019 of the 5 largest open DNS resolvers \cite{radu2020consolidation}, which constitutes 50.64\% of the total usage of global DNS. We test those DNS providers and obtained that 72.51\% of the users of those resolvers use open DNS resolvers that enable ECS. 
We analyzed the segmentation of the 10 largest public DNS providers and ISPs in the world, were together makeup 60.65\% of global DNS use. Using the RIPE ATLAS machines \cite{ripe_atlas} we were able to test 8 of them \footnote{we could not find machines in Century Link and Pakistan Telekom ISPs, both together holds 1.75\% of the DNS usage in the world.}, and we found that all the largest ISPs (Liberty Global, Comcast, Nevalink, Claro S.A., Korea Telecom, Telekom Austria) enable ECS. The only large open resolver that does not support ECS is Cloudflare \cite{cloudflare}, which does not provide it due to privacy reasons of exposing the user IP. In our solution, we propose to use a constant ECS IP address for each country so the device's IP will not be exposed.\\
To support resolvers that do not forward ECS (such as Cloudflare), we suggest to configure the authoritative name server to start an applicative connection (e.g., HTTP). That way, the IoT device will transfer his registered location. Our proposal does not require backward compatibility of the IoT devices, only an update of the vendors' server's side.

\begin{table}
\begin{tabular}{|l|c|c|l|}
\hline
\multicolumn{1}{|c|}{\textbf{DNS Provider}} & \textbf{ECS Enable} & \textbf{ ECS Forward} & \textbf{Share in 2019} \\ \hline
Google                                      & Yes                 & Yes                         & 35.94\%                \\
Quad9                                       & Yes                 & Yes                         & 0.78\%                 \\
Cloudflare                                  & No                  & No                          & 13.80\%                \\
OpenDNS                                     & No                  & No                          & 0.03\%                 \\
Yandex DNS                                  & No                  & No                          & 0.09\%                 \\
Comodo                                      & Yes                 & Yes                         & Unknown                \\
Verisign                                    & Yes                 & Yes                         & Unknown                \\
Alternate                                   & Yes                 & Yes                         & Unknown                \\
AdGuard                                     & Yes                 & Yes                         & Unknown                \\
UncensoredDNS                               & Yes                 & Yes                         & Unknown                \\
UltraRecursive                              & Yes                 & Yes                         & Unknown                \\
DNS.Watch                                   & Yes                 & Yes                         & Unknown                \\
Neustar                                     & Yes                 & Yes                         & Unknown                \\ \hline
\end{tabular}
\caption{ECS enable status of the 13 major public DNS resolvers and whether they perform forward ECS from the client.}
\label{table_ecs_enable_open_dns}
\end{table}

ECS is intended to be added by the resolver, but it can also be added by the end-user device (IoT in our case), although there is no reason mentioned in the RFC why such a situation will happen, as we can see in the RFC, it is certainly not inevitable: `'The ECS option should generally be added by recursive resolvers when querying authoritative name servers, as described in Section 12.  The option can \emph{also be initialized by a Stub Resolver} or Forwarding Resolver.'`
We used the Open Observatory of Network Interference dataset and indicates for the 13 major public DNS resolvers whether they enable ECS and whether they perform forward when the ECS is sent from the end-user. Other papers measured the ECS adoption, in our measurements, we analyze the ECS forward functionality. For all the public DNS resolvers we tested, we got answers for the ECS we sent, as presented in Table \ref{table_ecs_enable_open_dns} all the public DNS resolvers also perform forward ECS.

\begin{figure}
    \centering
    \includegraphics[width=0.96\linewidth]{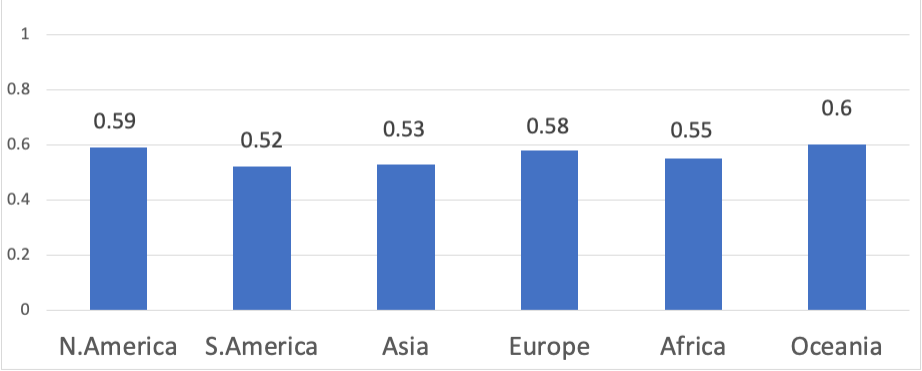}
    \caption[width=0.9\linewidth]{Adoption of ECS by continent segmentation.}
    \label{fig:cdn_exp}
\end{figure}

To measure the ECS adoption in another environment, we used 8,019 random machines in RIPE ATLAS \footnote{RIPE Atlas \cite{ripe_atlas} is a global, open, distributed Internet measurement platform, consisting of thousands of measurement devices that measure Internet connectivity in real-time.} and checked for each one whether the recursive resolver to which it is configured allows ECS or not. Our results show that 56.7\% of the machines returned to us with an ECS response and 43.3\% of the machines returned without ECS.
We examined whether there are differences in the segmentation of the results per continent, as can be seen in Figure \ref{fig:cdn_exp}, the distribution is widely uniform concerning the total result.

Al-Dalky et al.\cite{al2019look} analyze ECS deployment by recursive resolvers via passive observations from a large CDN perspective in the Passive Major CDN DNS traffic dataset. Over 3.7M resolvers, only 7,737 were sent at least one ECS query, that's only 0.2\% of the resolvers in the dataset. Public resolvers providers add resolvers around the world and improve the end-user location approximation, we assume that this is the reason that ECS is uncommon.
\begin{figure}
\begin{minipage}{0.9\linewidth}
\centering
\subfloat[Xiaomi light bulb]{\label{fig:xiaomi_equality_vs_naïve}    \includegraphics[width=1.1\linewidth]{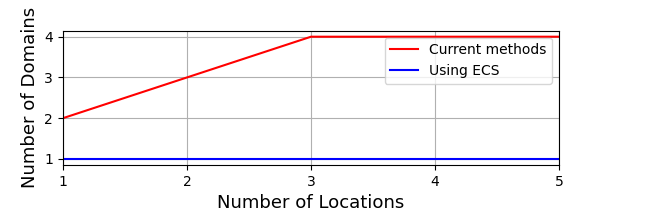}}
\end{minipage}%
\vskip 1.2cm
\begin{minipage}{0.95\linewidth}
\centering
\subfloat[Yi camera]{\label{fig:yi_equality_vs_naïve}    \includegraphics[width=0.96\linewidth]{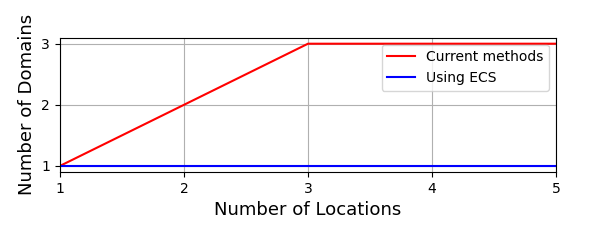}
}
\end{minipage}
\par\medskip

\caption{Comparison of ECS enabled MUD VS naïve unifying MUD files  Each point on the x-axis corresponds to the unified or ECS MUD at the specified number of locations. We ordered the locations according to places that are further away from each other (cross-regions), to gain more information in each iteration.} 
\label{fig:nainve_mud_vs_ecs_comparison}
\end{figure}  

\section{Case Study: Allowlists in the Manafacturer Usage Description (MUD) Framework}
\label{sec:MUD}

In previous sections, we present how user-defined location affects the device's network behavior (i.e., the domain names that the device connects to).
The set of domain names used by IoT devices directly affects IoT security tools that use this set to detect malicious traffic. Specifically, the IETF Manufacturer Usage Description (MUD) framework, generates allow-lists according to this set. 
In this section, we present a case study of using our ECS proposal with MUD. We show how by using user-defined location in the ECS field, the number of domains in the MUD allow-list is significantly reduced and the entire MUD's management and deployment processes become more streamlined. 
\label{sec:mud_archs}
\subsection{MUD Background}
MUD is an Internet standard~\cite{mud_ietf} that aims to reduce the attack surface for IoT devices by describing their appropriate traffic patterns. Any traffic that does not comply with this description is considered malicious and can be, for example, blocked. These descriptions are provided by the IoT manufacturers in \emph{MUD files}. 

MUD files consist of Access Control Lists (ACLs), each with several Access Control Entries (ACEs). Each ACE is defined as a 5-tuple: 
\vspace{-0.5em}
\begin{equation}
    \label{def:ACE_params}
    \begin{split}
        ACE = (legitimate\_endpoints,protocol,source\_port, \\
       destination\_port, direction)
    \end{split}
\end{equation}

The legitimate endpoints are the endpoints with which the IoT connects they are commonly defined by domain name, IP, or MAC for intra-LAN scenarios. 
Note that the MUD RFC highly recommends avoiding the use of IP addresses and encouraging the use of domains instead.

The corresponding action of the ACE is typically to either “accept”  or “drop”. Because the MUD file specifies an allow-list, the default rule is to drop traffic that does not correspond to any ACE.

The MUD framework itself consists of several components. A \emph{MUD manager}, also known as the \emph{MUD controller}, is responsible for obtaining and processing the MUD information. For each IoT device, the MUD manager first obtains the MUD file from its manufacturer's \emph{MUD server}. The MUD server's address for the IoT device is stored as a \emph{MUD URI} in the device's firmware. This URI can be obtained by the MUD manager in a variety of ways as specified in the RFC. Nevertheless, it is most commonly obtained through a dedicated option in the DHCP protocol, which the IoT device executes to connect to the network. 
With the MUD file at hand, the MUD manager parses the file and installs the corresponding ACL rules on a network security device, such as a firewall or AAA server, to reduce the attack surface on the device.

In current MUD architectures, the MUD manager fetches MUD files from the MUD file server. The MUD manager is not aware of the user-defined location since it is configured by the user on the endpoint device. If the vendors would use an IP-based location to identify the IoT location, the implementation of MUD in a network can be straightforward; using Geo-IP a relevant MUD would be fetched and applied. But then the user will have no choice of the device location. This unexpected requirement of IoT vendors raises problems when trying to apply MUD in networks. On one hand, the MUD profile must be adapted to the user-defined location, and on the other hand, the user-defined location is determined by the user and can be changed as of the user's decision.
One can suggest a 'trivial' solution, in which the manufacturer maintains a single MUD file, covering all the networking flows of all available locations.
This proposal requires high maintenance and many different domains across locations are needed. 
Manufacturers are faced with the challenging task of creating a comprehensive and representative MUD that takes into account many parameters. To overcome these challenges, some tools generate MUD files from network captures such as MUDgee and MUD-PD \cite{mudgee_paper,MUD_PD}.

\subsection{MUD using ECS}
When using ECS the manufacturer maintains a single shorter MUD file. The device publishes the MUD-URL and the MUD manager fetches it. As depicted in Figure \ref{fig:ecs_basic_vs_user_defined_b}, this ECS architecture reduces the use of several domains to distinguish between locations. Instead, the device adds an ECS field to its DNS requests and gets its (user-defined) local server address. The ECS parameter holds a user-defined location and not the real IP of the device. It allows the device to set it dynamically, to support a user-defined location and not an IP-based one.
Manufacturers currently use several domains (i.e., see Figure \ref{fig:xiaomi_germany_china}) to separate servers of different user-defined location regions. Using ECS, domain identifiers separation is unnecessary, it 'shifts left' the separation into the network level. This network-assisted MUD suggestion leverages the ECS option to gain flexibility without introducing any disruptive changes for the manufacturer. Note that there are several MUD architectures proposed by NIST \cite{nist_mud_archs} in which the MUD manager installs the corresponding ACL rules by intercepting DNS requests or by issuing DNS requests \cite{ietf-opsawg-mud-iot-dns-considerations-05}. When the MUD manager issues DNS requests, it should send them with no ECS configured. Then, the authoritative replies with a list of all the corresponding IPs (of all the available zones). The location separation is performed using ECS, reducing the number of rules that are enforced by a security gateway. It also reduces the cost of maintaining several domains, MUD files, and other applicative needs. ECS resolves to different servers in a more cost-effective way than using several domain identifiers.
\subsection{Performance Evaluation}
To evaluate current solutions, we examined two devices: Xiaomi Light bulb and YI Camera, each captured in 10 locations. Using MUDgee, we created a single MUD file for each user-defined location, covering the network behavior in these locations. 
To create a single MUD file that covers all the potential flows, we unified the MUD files from all available user-defined locations and created a single unified MUD for each device.
Then, we calculate the number of required domains when using ECS to distinguish user-defined locations and while using several domains, each for each location.
Figure \ref{fig:nainve_mud_vs_ecs_comparison} presents a comparison of the naive unified MUD, to our ECS proposal in regards to the number of domains. The number of domains required in current methods is higher by more than 66\% than our proposal.

\begin{figure*}
\begin{minipage}{.6\linewidth}
\centering
\subfloat[Yi Camera in Mexico and UK]{\includegraphics[width=0.96\linewidth]{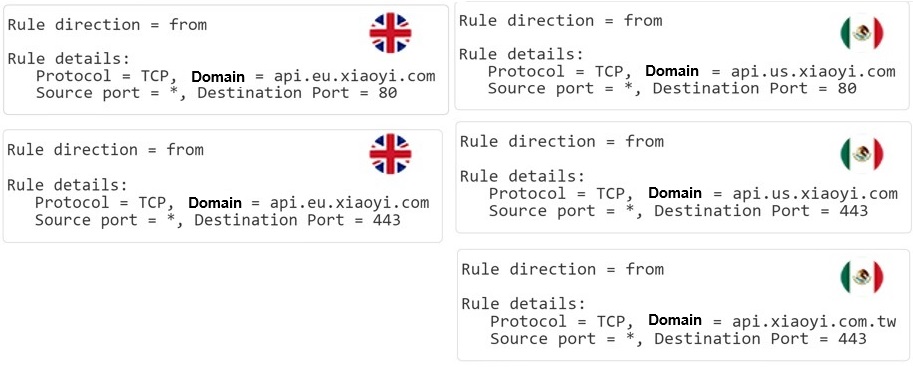}}
\end{minipage}%
\begin{minipage}{.3\linewidth}
\centering
\subfloat[Yi Camera: MUD Using ECS]{    \includegraphics[width=0.96\linewidth]{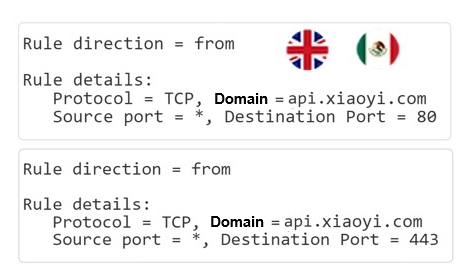}
}
\end{minipage}\par\medskip

\caption{MUD files of YI camera MUDs in UK (left,a) and Mexico (right,b).} 
\label{fig:generalization_mud_example}
\end{figure*}

Figure \ref{fig:generalization_mud_example} presents the MUD files of the YI Camera in two user-defined locations and their ECS-enabled MUD.

\section{Related Work}
To the extent of our knowledge, this is the first work that defines IoT device location as a factor that impacts a device's network behavior. The only related work we are aware of deals with the influence of privacy regulations (GPDR, FTC) on the network behavior of IoT in the United Kingdom and the United States \cite{ren2019information}. In contrast, our work investigates the impact of location in many different countries and demonstrates that there exist other reasons for the differences, such as cloud regions, marketing motivations, and more.

Several studies have investigated ECS from several perspectives. 
In~\cite{al2019look} the authors look at the ECS-related behavior of recursive resolvers and some ECS implications for DNS caching, they analyze ECS deployment by recursive resolvers via passive observations from a large CDN perspective and “in the wild”. From the passive observations, with over 3.7M resolvers, only 0.2\% were sent at least one ECS query.
In our work we analyze the ECS deployment in RIPE ATLAS machines, those machines are located all around the globe with different DNS providers (open DNS and ISPs).
In~\cite{radu2020consolidation} the authors present the top DNS resolver providers, they show that more than 50\% of the world use open resolvers, we used these results and check by RIPE ATLAS machines if those DNS resolvers are enabling ECS. We obtained that 86.2\% of the ISPs and open resolvers enable ECS and 72.51\% of the largest DNS open resolvers in the world enable ECS.
Vries et al. \cite{de2019passive} use 2.5 years of passive ECS-enabled queries to study Google Public DNS. The authors show that DNS traffic to Google Public DNS is frequently routed to data centers outside the country even though a local data center is available in-country. We used their conclusions to demonstrate the problem in associating Resolvers with the end-user location using the Anycast approach, and as an explanation for creating the ECS.

\bibliographystyle{IEEEtran}
\bibliography{references}
\end{document}